\documentclass[fancyheadings,twoside,psfig,float,fleqn,epsf,12pt]{article}
\usepackage{psfig}
\usepackage{epsf}
\usepackage{amssymb}
\usepackage{float}
\usepackage{fancyheadings}

\restylefloat{figure}

\renewcommand{\theequation}{\arabic{section}.\arabic{equation}}

\newcommand{\be}{\begin{equation}}
\newcommand{\ee}{\end{equation}}
\newcommand{\bea}{\begin{eqnarray}}
\newcommand{\eea}{\end{eqnarray}}
\newcommand{\bean}{\begin{eqnarray*}}
\newcommand{\eean}{\end{eqnarray*}}
\newcommand{\finishproof}{${{}_\blacksquare}$}

\newtheorem{lemma}{Lemma}[section]

\newtheorem{conjecture}{Conjecture}[section]

\newenvironment{proof}{{\flushleft \bf Proof:}}{}

\newenvironment{acknowledgment}{{\flushleft \bf Acknowledgment:}}{}

\def\tink{$^\prime$}       
\def\addit{%
    \let\oldequation=\theequation
    \def\theequation{\oldequation\tink}%
}
\def\endit{%
    \let\theequation=\oldequation
}

\begin{document}

\title{
Dissipative Behavior of Some Fully Non-Linear KdV-Type Equations
}
\author{Yann Brenier\footnotemark[2] \and Doron Levy\footnotemark[3]}
\date{}

\renewcommand{\thefootnote}{\fnsymbol{footnote}}
\footnotetext[2]{Universit\'{e} Pierre et Marie Curie, 4 Place Jussieu, 75252 Paris Cedex 05, France;
Email: {\tt brenier@dmi.ens.fr}}
\footnotetext[3]
 {Department of Mathematics, University of California, Berkeley, CA 94720, 
  and Lawrence Berkeley National Lab; {\tt dlevy@math.berkeley.edu}}
\renewcommand{\thefootnote}{\arabic{footnote}}

\maketitle

\begin{abstract}
The KdV equation can be considered as a special case of the general equation
\be
 u_{t} + f(u)_{x} - \delta g(u_{xx})_x = 0,  \qquad \delta > 0,  
\label{eq:model.int}
\ee
where $f$ is non-linear and $g$ is linear, namely $f(u)=u^2/2$ and $g(v)=v$. 
As the parameter $\delta$ tends to $0$, 
the dispersive behavior of the KdV equation has
been throughly investigated (see, e.g.,  
\cite{whitham}, \cite{lax-levermore}, \cite{drazing:solitons}
and the references therein).
We show through numerical evidence that a completely different, dissipative
behavior occurs when $g$ is non-linear, namely when $g$ is
an even concave function such as $g(v)=-|v|$ or $g(v)=-v^2$.
In particular, our numerical results hint that as 
$\delta \rightarrow 0$ the solutions
{\em strongly converge} to the unique entropy solution of the formal
limit equation, in total contrast with the solutions of the KdV equation.
\end{abstract}

\bigskip
\noindent
{\bf PACS.} 05.45.-a; 02.70.Bf.

\bigskip
\noindent
{\bf Key words.} KdV-type equations.  Finite-difference methods.  Nonlinear dynamics.


\section{Introduction}
\setcounter{equation}{0}\setcounter{figure}{0}

Let us consider the first-order non-linear evolution PDE
\be
  u_{t} + f(u)_{x} = 0,    \label{eq:scalar.conservation}
\ee
with a strictly convex non-linearity $f''(u) \geq \alpha > 0$, typically
$f(u)=u^2/2$ which corresponds to the usual inviscid Burgers equation
\cite{whitham}.
This equation generally produces shock waves in a finite time and
solutions must be understood in a suitable weak sense.
A good framework enforcing existence, uniqueness, $L^1$ stability,
(\cite{smoller}, \cite{leveque}, \cite{serre},...)
is provided by the so-called entropy solutions,
obtained through the vanishing viscosity
method by passing to the limit in the viscous approximation
\be
  u_{t} + f(u)_{x} - \delta u_{xx}=0,    \label{eq:viscous.approximation}
\ee
as $\delta>0$ tends to zero \cite{smoller}.
Entropy solutions can be also characterized as the weak solutions
of (\ref{eq:scalar.conservation}) which satisfy the Oleinik
one-sided Lipschitz condition (OSLC)
\be
  u_{x}(t,x)\le \frac{1}{\alpha t}.    \label{eq:oleinik}
\ee
A completely different behavior is observed if 
(\ref{eq:scalar.conservation}) is approximated by the KdV equation
\be
 u_{t} + f(u)_{x} - \delta u_{xxx} = 0,  \qquad \delta > 0,
\label{eq:kdv}
\ee
as investigated by many authors, in particular \cite{lax-levermore}.  
When the formal limit has shocks, the corresponding
solution of the KdV equation becomes highly oscillatory and does not converge
at all to the entropy solution (even in a weak sense). 

An alternative approach of approximating equations such as 
(\ref{eq:scalar.conservation}) is to design a numerical scheme.
In \cite{brenier-osher:oslc} Brenier and Osher presented a 
second-order accurate version of the Roe scheme 
(\cite{smoller}, \cite{leveque})
and an OSLC consistent method of lines
for approximating solutions of scalar conservation laws,
(\ref{eq:scalar.conservation}).
As was already pointed out in \cite{brenier-osher:oslc}, the ``modified 
equation'' of this scheme (that is to say the asymptotic PDE of the
discrete equation) is of interest since (as it is second-order) 
it must involve, just like the KdV equation, a third-order term,
but unlike the KdV equation, 
due to the enforced OSLC this third-order term must enforce 
restrictions on the possible creation of wiggles.  
Our present work presents an extended 
numerical study of these ``non-classical dispersive'' effects,
as well as an extended study of the numerics involved.

The model equations we consider are 
(\ref{eq:model.int}) where $g$ is concave and even,
typically $g(s)=-|s|$ and the smoother
$g(s)=-s^2$.
The numerical tools taken from \cite{brenier-osher:oslc}
enable us to approximate the solutions of (\ref{eq:model.int})
with a numerical method in which its modified equation takes the
desired form of 
\be
  u_{t} + f(u)_{x} - \delta g(u_{xx})_x = - \epsilon u_{xxxx} + LOT,   \label{eq:model.2.int}
\ee
with a positive $\epsilon \sim \delta \Delta x$, where $LOT$ means perturbation
terms with at most third-order space derivatives (low order terms).
In the specific case where $g(s)=-|s|$, there are no 
low order terms on the RHS of (\ref{eq:model.2.int})
and hence we are left only with a {\em linear\/} dissipative term 
proportional to $\delta \Delta x$, which enables us to hold a fixed $\delta$ while
eliminating the RHS at the limit $\epsilon \rightarrow 0$.

The analytical tools available to study fully non-liner
evolution PDEs such as 
(\ref{eq:model.int}) and (\ref{eq:model.2.int}),
seem to be limited and a rigorous theory is, in our opinion, a
very challenging task that deserves to be further addressed.
Our main results are based on numerical evidence.

The simple {\em non-linear switch\/} in the third-order term
$|u_{xx}|_x$ based on the sign of the second derivative $u_{xx}$,
has a major impact. The same holds true for other non-linear
third-order operators such as $(u_{xx}^2)_x$.
Instead of the usual dispersive wave-train that
develops in the KdV equation, here we still observe the appearance
of a {\em single \/} ripple, which indeed can be still considered as a 
dispersive phenomenon.  
This ripple, however, is of a totally different character than what we
are familiar with in the KdV equation.
Instead of developing into a full wave-train which propagates in time, 
this ripple remains {\em single\/} during the evolution of the solution.
It does not increase above a 
certain size which is related to the value of the parameter $\delta$
in (\ref{eq:model.int}). 

The striking numerical discovery is that not only this ripple is single, but
it also {\em disappears\/} in the limit $\delta \rightarrow 0$.
It turns that (\ref{eq:model.int}) is much closer in its behavior to the 
dissipative, viscous 
approximations of the form
\[
 u_{t} + f(u)_{x} = \epsilon u_{xx},   \qquad \epsilon > 0,
\]
than to the KdV equation
\[
 u_{t} + f(u)_{x} + \delta u_{xxx} = 0,
\]
for which at
the same limit $\delta \rightarrow 0$, the dispersive waves increase.

The natural conjecture that follows is that (\ref{eq:model.2.int}),
{\em strongly converges\/} to the
entropy solution of the scalar conservation law (\ref{eq:scalar.conservation}) as
$\delta$ and $\epsilon\delta^{-1}$ tend to $0$.

We would like to emphasize that when using numerical methods in order to 
approximate solutions of differential equations, instead of solving the original 
problem, we end up solving a modified problem, which typically includes high-order terms.  
Here, we are interested in studying the effects of such high-order terms in the
particular case in hand.  We believe that the properties of the model 
equations we study in this work, might provide insight for the inclusion of 
similar nonlinear terms in future models.

The paper is organized as follows:
we start in \S\ref{section:motivation} by presenting both our model equations and
the numerical method of approximating their solutions.
We continue in \S\ref{section:properties} by commenting on the major properties
of these equations.  In particular, we study traveling waves
solutions and state a formal one-sided bound (OSLC-type) on the derivative.

We then continue in \S\ref{section:strong} by presenting several numerical 
examples which leads us to our main conjecture stated above, which is formalized
in Conjecture \ref{conjecture:strong}.

We end in \S\ref{section:spurious} by demonstrating and analyzing spurious 
numerical oscillations that appear when solving our model problems with a certain
choice of parameters.  
The methods we use for analyzing the binary oscillations follow
Goodman and Lax, \cite{goodman-lax:dispersive}.  Similar techniques were used, e.g.,
in \cite{hayes-97, hayes-98, levermore-liu}.
This study is aimed at gaining a further understanding in
order to be able to distinguish between the properties of the equations and the
appearing numerical phenomena which depend on the specific method used when approximating
the solutions.

\begin{acknowledgment}
The research was supported in part by TMR grant \#ERBFMRXCT960033.  
The work of D.L.\ was supported in part by the Applied Mathematical Sciences 
subprogram of the Office of Science, U.S.\ Department of Energy, under 
contract DE--AC03--76--SF00098.  Part of the research was done while
D.L.\ was affiliated with ENS Paris.  D.L. would like to thank E.\ Tadmor 
and A.\ Kurganov for stimulating discussions.
\end{acknowledgment}

%
%
\section{Motivation}    \label{section:motivation}
\setcounter{equation}{0}
\setcounter{figure}{0}

\subsection{The Model Equations}      \label{subsection:model}

The motivation for our study is 
the modified equation of the Brenier-Osher (BO) scheme presented in \cite{brenier-osher:oslc}.
We start by considering the scalar conservation law (\ref{eq:scalar.conservation})
with a (strictly) convex flux function, $ f''(u) \geq \alpha > 0$.

We discretize the space using a fixed mesh spacing, $\Delta x$, denoting the approximation
of the solution at the discrete grid-points by $u_{i} := u(i \Delta x)$.

The BO method for approximating solutions of (\ref{eq:scalar.conservation}) is a 
second-order convergent Roe scheme and an OSLC consistent method of lines, which can be
written as
\be
\frac{d u_{i}}{dt} + \frac{1}{\Delta x}(f_{i+1/2}-f_{i-1/2}) = 0,  \label{eq:bo}
\ee
with
\[
  f_{i+1/2} = h \left( u_{i} + \frac{\Delta x}{2}s_{i}, u_{i+1} - \frac{\Delta x}{2} s_{i+1} \right).
\]
Here, $h$ is the Godunov flux associated with $f$,
\[
 h(a,b) = h_{\mbox{\tiny  GOD}}(a,b) = \epsilon \min_{0 \leq s \leq 1} \epsilon f(a+s(b-a)), \quad
   \epsilon = sgn(b-a),
\]
and the slopes, $s_{i}$, are limited by
\[
 s_{i} = \frac{1}{\Delta x}\max(u_{i+1}-u_{i},u_{i}-u_{i-1}).
\]

We note that the 
OSLC consistency of (\ref{eq:bo}) is in the sense that the discrete approximation satisfies
\[
 \frac{u_{i+1}(t)-u_{i}(t)}{\Delta x} \leq \frac{1}{p(u(0,\cdot))^{-1} + \alpha t},
\]
where $p(v)$ is the one-sided Lipschitz semi-norm,
\[
 p(v) = \sup_{x \ne y} \left( \frac{v(x)-v(y)}{x-y} \right)_{+}, \qquad a_{+} = \max(0,a).
\]

In the linear case, i.e. for $f(u)= \beta u$ with $\beta$ constant, the BO scheme 
yields the modified equation
\be
 u_t + f(u)_{x} + \delta |u_{xx}|_{x} = 0,   \qquad \delta = O(\Delta x^2).  \label{eq:mod}
\ee
In the nonlinear case for a general $f(u)$, additional low-order terms appear in (\ref{eq:mod}).
The existence of these terms was ignored in \cite{brenier-osher:oslc}.

As was already pointed out in \cite{brenier-osher:oslc}, this modified equation (\ref{eq:mod})
is of interest since it involves a dispersion term (as it is second-order), but unlike KdV, due
to the enforced OSLC, this third-order term does not create wiggles.  These are the properties
this work is aimed at studying.

Since (\ref{eq:mod}) includes a non-smooth flux, we extend the current discussion to
a wider family 
\be
 u_t + f(u)_{x} - \delta g(u_{xx})_{x} = 0,    \label{eq:mod.2}
\ee
where $g(u_{xx})$ is assumed to be concave, $g''(s) < 0, \forall s$.  There are no further 
regularity assumptions on $g$.   

To summarize, our model equations are (\ref{eq:mod}) and its smooth version (\ref{eq:mod.2}).

\subsection{The Numerical Scheme}    \label{subsection:numerical}
For the following analysis the precise derivation and the origin of (\ref{eq:mod}) and (\ref{eq:mod.2})
is no longer important.  It is their properties that we want to study.
To proceed, we write a discretization of (\ref{eq:mod}),(\ref{eq:mod.2}).  
It is remarkable that it is possible to write down such a discretization whose
modified equation involved a {\em linear} fourth-order derivative of the ``correct'' 
(dissipative) sign.

Following \cite{brenier-osher:oslc}, a straightforward discretization of the generalized
modified equation (\ref{eq:mod.2}) will be, e.g., to define the (first-order) numerical flux
$g_{i+1/2} = g(w_i,w_{i+1})$ as the Engquist-Osher (EO) flux
\be
 g_{\mbox{\tiny EO}}(a,b) = g(b_{+})+g(a_{-}), \qquad b_{+} = \max(0,b), \quad a_{-} = \min(0,a).
\ee
A method of lines for (\ref{eq:mod.2}) can be then written as
\be
 \Delta x \frac{d u_{i}}{dt} + \Delta_{-} \left[ f\left( \frac{u_i+u_{i+1}}{2} \right) - 
    \delta g(w_i,w_{i+1}) \right] = 0,   \label{eq:mod.2.lines}
\ee
where $\Delta_{-}a_{i} = a_i - a_{i-1}$, and $w_{i} = (u_{i-1} - 2u_i+u_{i+1})/(\Delta x)^2$.

Replacing the time derivative in (\ref{eq:mod.2.lines}) by the forward-Euler discretization
associated with a fixed time-step, $\Delta t$, we obtain the fully-discrete
\be
 \frac{u_{i}^{n+1}-u_{i}^{n}}{\Delta t} + 
  \frac{1}{\Delta x} \Delta_{-} \left[ f\left(\frac{u_{i}^{n}+u_{i+1}^{n}}{2}\right)
    - \delta g(w_{i}^{n},w_{i+1}^{n}) \right] = 0,   \label{eq:mod.2.discrete}
\ee
where for the space-time discretization we use the notation $u_{i}^{n} := u(i \Delta x, n \Delta t)$.

A somewhat tedious but straightforward computation shows that the limiting equation for the
fully-discrete process (\ref{eq:mod.2.discrete}) has the form of
\bea
 \lefteqn{u_t + f(u)_{x} - \delta g(u_{xx})_{x} =} \label{eq:discrete.mod} \\
 && = \Delta x \delta \left[ \frac{1}{2} sgn(u_{xx}) g'(u_{xx}) u_{xxxx} + 
   \frac{1}{2} sgn(u_{xx}) g''(u_{xx}) u_{xxx}^{2} \right] + O((\Delta x)^2,\Delta t). \nonumber
\eea
In particular, for the choice $g(s) = -|s|$, (\ref{eq:discrete.mod}) becomes
\be
 u_t + f(u)_{x} + \delta |u_{xx}|_{x} = -\epsilon u_{xxxx}, \qquad \epsilon = \frac{1}{2}\Delta x \delta.
   \label{eq:abs.mod}
\ee
For the more regular choice of $g(s) = -s^2$, (\ref{eq:discrete.mod}) takes the form of
\be
 u_t + f(u)_{x} + \delta (u_{xx}^{2})_{x} = - \Delta x \delta \left[ |u_{xx}|u_{xxxx} 
   + \frac{1}{2} sgn(u_{xx}) u_{xxx}^{2} \right].   \label{eq:power.mod}
\ee

At that point, we add equations (\ref{eq:abs.mod}),(\ref{eq:power.mod}) to our two
previous model equations, (\ref{eq:mod}) and (\ref{eq:mod.2}).
%
%
\section{Properties of the Model Equation}    \label{section:properties}
\setcounter{equation}{0}
\setcounter{figure}{0}

In this section we discuss some of the major properties of (\ref{eq:mod.2}).
We will assume that $f(u)$ is convex.  For demonstration purposes if will
be usually taken as $f(u) = u^{2}/2$.
We also assume a concave $g(s)$, which will be either $-|s|$ or $-s^2$,
unless mentioned otherwise.
Namely, we deal with two model equations.  The first equation
corresponds to the choice of $g(s)=-|s|$, 
\be
  u_{t} + u u_x + \delta |u_{xx}|_{x} = 0.   \label{eq:example.abs}
\ee
The second choice corresponds to the smoother, $g(s)=-s^{2}$, 
\be
  u_{t} + u u_x + \delta (u_{xx}^{2})_{x} = 0.    \label{eq:ex.smooth}
\ee  

We note that in both equations (\ref{eq:example.abs}) and (\ref{eq:ex.smooth}),
one can eliminate by rescaling the parameter $\delta$.
This is done in (\ref{eq:example.abs}) by the rescaling $u \rightarrow \delta u$ and
$t \rightarrow \delta t$.  Such a rescaling also holds for (\ref{eq:abs.mod}) when
$\epsilon \sim \delta \Delta x$.
The different scaling $x \rightarrow \delta^{-1/4}x$ and $t \rightarrow \delta^{-1/4}t$
gets rid of $\delta$ in (\ref{eq:ex.smooth}).

We would also like to note that $u_{t}+uu_{x}+\delta g(u_{xx})_{x} = - \epsilon u_{xxxx}$
is invariant under the Galilean transformation $u(x,t) \rightarrow u(x+ct,t)$ and
$u \rightarrow u-c$.

\subsection{Traveling Waves}   \label{subsection:traveling}

We compute a monotonic profile of the traveling waves solution for (\ref{eq:ex.smooth}).
The monotonicity is an important property of these traveling waves.
Our numerical experiments below, indicate that these particular traveling wave solutions behave
as attractors.

We impose a symmetric upstream-downstream profile with 
$u_{\mbox{\tiny Left}} = - u_{\mbox{\tiny Right}} = u_{1} > 0$.
One integration of (\ref{eq:ex.smooth}) results with 
\[
 \frac{u^2}{2} - \frac{u_{1}^{2}}{2} = - \delta u_{xx}^{2},
\]
which can be integrated twice to yield
\be
 x = - (2 \delta)^{1/4} \int_{0}^{u}\left(\pm \left(s\sqrt{u_{1}^{2}-s^{2}}+
   u_{1}^{2} \arcsin\left(\frac{s}{u_{1}} \right) - \frac{\pi}{2}u_{1}^{2} \right)\right)^{-1/2}ds.  
   \label{eq:sol.trav}
\ee
The sign in the integrand is set according to the sign of $u$.
In particular, for $u_{1}=1$, (\ref{eq:sol.trav}) becomes
\be
 x = - (2 \delta)^{1/4} \int_{0}^{u}\left(\pm \left(s\sqrt{1-s^{2}}+
   \arcsin s - \frac{\pi}{2} \right)\right)^{-1/2}ds.  \label{eq:sol.trav.1}
\ee

We demonstrate this profile in the following example in which
we approximate solutions of (\ref{eq:ex.smooth}) using the
numerical scheme presented in \S\ref{subsection:numerical}.
Here, we used Riemann initial data
as
\be
 u_{0}(x) = \left\{
   \begin{array}{rl}
	1 & x < 0 \\
	-1 & x > 0. 
   \end{array}
 \right.          \label{eq:example.riemann}
\ee

\begin{figure}[H]    
\begin{center}
  \leavevmode {
  \hbox{
        \epsfxsize=10cm
        \epsffile{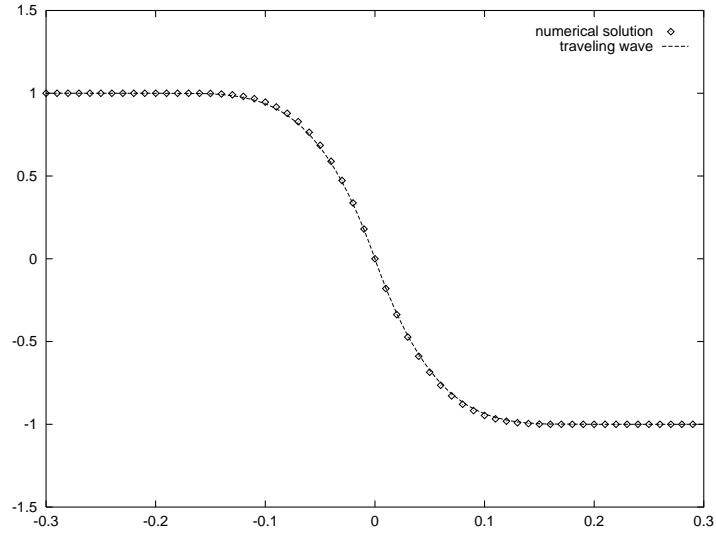}
       }
  }
\end{center}
\caption{{\sf Convergence of the Numerical Solution to the Traveling Wave Solution, 
              $\delta=10^{-5}$, $T=0.5$} \label{figure:trav.1}}
\end{figure}

\begin{figure}[H]    
\begin{center}
  \leavevmode {
  \hbox{
        \epsfxsize=10cm
        \epsffile{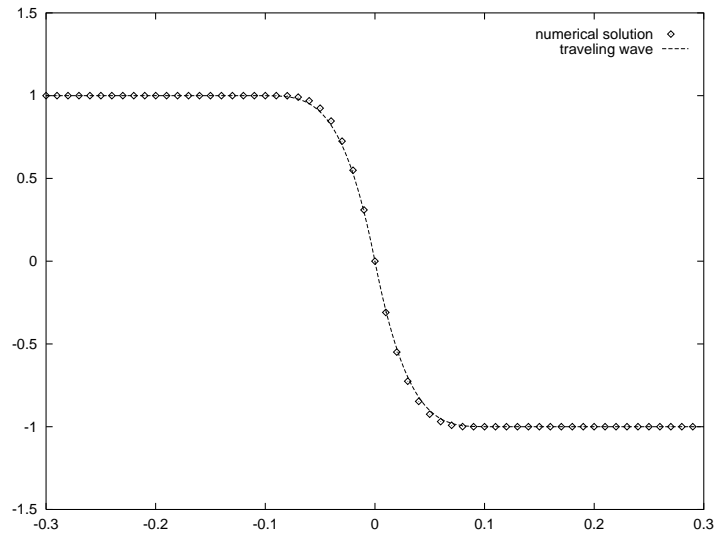}
       }
  }
\end{center}
\caption{{\sf Convergence of the Numerical Solution to the Traveling Wave Solution, 
            $\delta=10^{-6}$, $T=0.5$} \label{figure:trav.2}}
\end{figure}

The results are plotted in figures \ref{figure:trav.1}--\ref{figure:trav.2}
for different values of $\delta$, after the steady-state was reached.
These numerical results hint 
that the traveling wave solutions behave as attractors to the approximate solutions.

\subsection{The One-Sided Lipschitz Condition}    \label{subsection:oslc}
It is surprisingly easy to derive a formal one-sided
bound on the derivative $u_{x}$ in (\ref{eq:mod.2}).  Here, we have to augment (\ref{eq:mod.2})
with additional assumptions on $g(s)$ given in the following elementary lemma 

\begin{lemma}[OSLC]   \label{lemma:oslc}
Consider (\ref{eq:mod.2}).  Assume that $\forall s$, $f''(s)>0, g''(s)<0$ and that $g'(0)=0$.
Then the derivative, $u_x$, satisfies the formal one-sided estimate
\be
 \max_{x} u_{x}(t) \leq \max_{x} u_{x}(0), \quad \forall t>0.      \label{eq:oslc}
\ee
\end{lemma}
\begin{proof}
 Differentiating (\ref{eq:mod.2}) once yields
\be
 u_{tx} + (f'(u)u_{x})_{x} - \delta(g'(u_{xx})u_{xxx})_{x} = 0.  \label{eq:oslc.a}
\ee
Denoting $w=u_{x}$, (\ref{eq:oslc.a}) can be rewritten as
\[
 w_t + f''(u)w^{2}+f'(u)w_{x}-\delta g''(w_{x})w_{xx}^{2} - \delta g'(w_x)w_{xxx}=0.
\]
If we now take $v(t) = \max_{x}w(x,t)$, 
then due to the assumptions of the lemma, $\dot{v}<0$ and (\ref{eq:oslc})
follows.  \finishproof
\end{proof}

The appearance of a fourth-order derivative in (\ref{eq:discrete.mod})--(\ref{eq:power.mod})
prevents us from writing an analog of lemma \ref{lemma:oslc} in this case.  The fact that
the fourth-order derivative appears in the ``correct'' negative sign in the RHS of the equations
is associated with a dissipative process, but to quantify this in terms of the desired bound
on the derivative seems to be a highly non-trivial task.

%
%
\section{Strong Convergence?}        \label{section:strong}
\setcounter{equation}{0}
\setcounter{figure}{0}

In this section we numerically study the limit of solutions of (\ref{eq:mod})--(\ref{eq:mod.2})
as the parameter $\delta$ tends to zero.
Our starting point is the results obtained by approximating solutions of (\ref{eq:example.abs})
subject to the exponentially decaying initial conditions
$u(x,0) = e^{-100 x^{2}}$.
Similar phenomena to those we describe below, appear with other choices of initial data.

\begin{figure}[H]    
\begin{center}
  \leavevmode {
  \hbox{
        \epsfxsize=12cm
        \epsffile{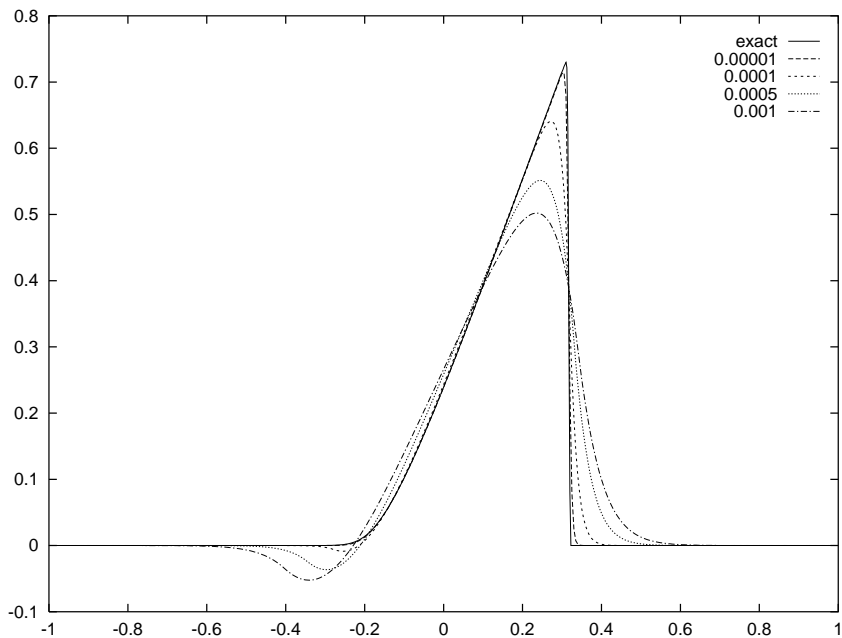}
       }
  }
\end{center}
\caption{{\sf Absolute-Value Dispersion with Exponential Initial Data.  
              $N=400$,  $T=0.5$, $\Delta t \cdot \delta = 10^{-8}$.  
              The different plots refer to different values of $\delta$.
	The ``exact'' curve is the entropy solution of the Burgers
	equation subject to the same initial data} \label{figure:ex1.1}}
\end{figure}

Figure \ref{figure:ex1.1} presents our most interesting discovery.
One can observe the creation of a {\em single\/} dispersive ``ripple'' which
{\em disappears\/} as $\delta \rightarrow 0$.
This phenomenon is totally different from what we know about solutions of
dispersive equations (KdV-type).  There,
an infinite dispersive wave-train develops in time 
(compare, e.g., with \cite[p. 64]{drazing:solitons}).  This wave-train increases
at the same limit $\delta \rightarrow 0$.
In that limit, the solution appears to strongly converge to the 
entropy solution of the Burgers equation subject to the same initial data.

\begin{figure}[H]    
\begin{center}
  \leavevmode {
  \hbox{
        \epsfxsize=12cm
        \epsffile{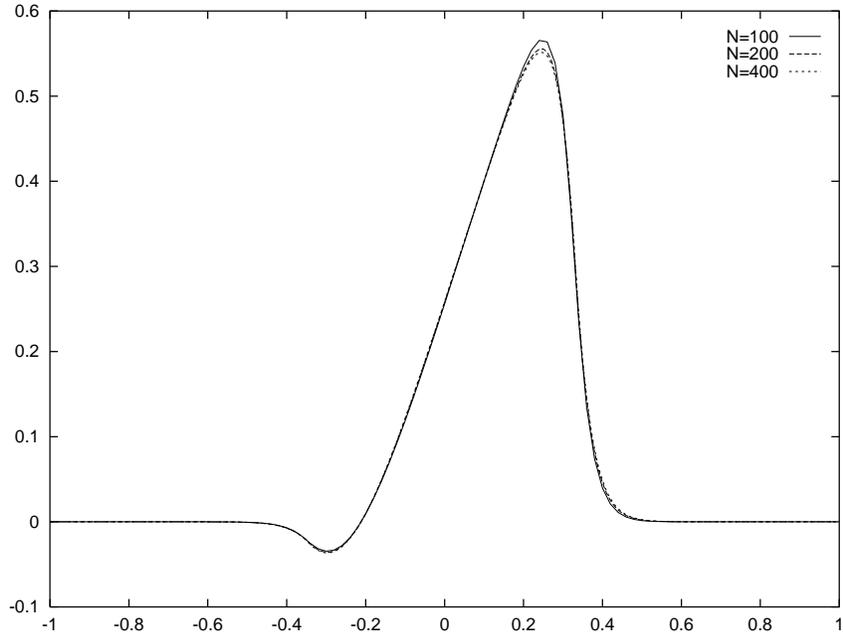}
       }
  }
\end{center}
\caption{{\sf Absolute-Value Dispersion with Exponential Initial Data.
              $\delta=5 \cdot 10^{-4}$,  $T=0.5$.  
		For $N=400$, $\Delta t=2 \cdot 10^{-5}$.
		For $N=200$, $\Delta t=2 \cdot 10^{-4}$.
		For $N=100$, $\Delta t=2 \cdot 10^{-3}$} \label{figure:ex1.2}}
\end{figure}

We confirm that the results shown in figure \ref{figure:ex1.1} are independent of
the spatial spacing $\Delta x$, by demonstrating in figure \ref{figure:ex1.2} 
a case in which $\delta$ is being held fixed while approaching the limit
$\Delta x \rightarrow 0$.

\begin{figure}[H]    
\begin{center}
  \leavevmode {
  \hbox{
        \epsfxsize=12cm
        \epsffile{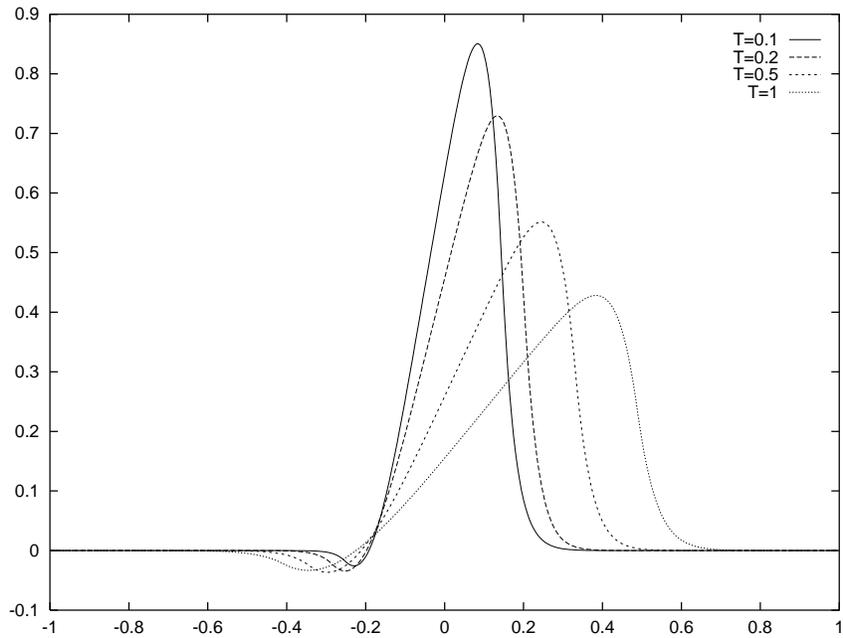}
       }
  }
\end{center}
\caption{{\sf Absolute-Value Dispersion with Exponential Initial Data.  
	      Time evolution of the solution.
              $N=400$, $\delta=5 \cdot 10^{-4}$, $\Delta t=2\cdot 10^{-5}$} \label{figure:ex1.3}}
\end{figure}

The evolution in time of the solution is shown in figure \ref{figure:ex1.3}.
We note that the dispersive ripple does not increase in magnitude above a certain
size (which by figure \ref{figure:ex1.1} is determined by the value of $\delta$).
No other ripples are being developed.

\begin{figure}[H]    
\begin{center}
  \leavevmode {
  \hbox{
        \epsfxsize=12cm
        \epsffile{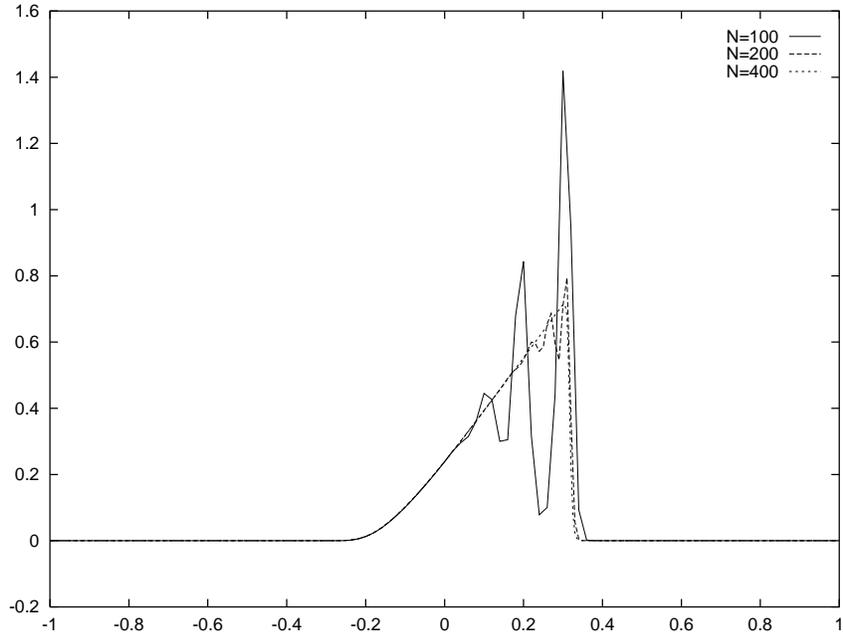}
       }
  }
\end{center}
\caption{{\sf Absolute-Value Dispersion with Exponential Initial Data.  
	      The appearance of numerical oscillations for small   
              $\delta=10^{-5}$.  $T=0.5$.  For $N=200$, $\Delta t=10^{-4}$.
              For $N=100$, $\Delta t=10^{-3}$} \label{figure:ex1.4}}
\end{figure}

In figure \ref{figure:ex1.4} we present a fully numerical phenomenon in which
spurious oscillations propagate from the shock.  This happens when the equation
is numerically solved for a relatively small value of $\delta$ and a relatively
large $\Delta x$.  We observe that these oscillations disappear as $\Delta x \rightarrow 0$.
A further study of similar numerical phenomena is presented in \S\ref{section:spurious}.

\begin{figure}[H]    
\begin{center}
  \leavevmode {
  \hbox{
        \epsfxsize=12cm
        \epsffile{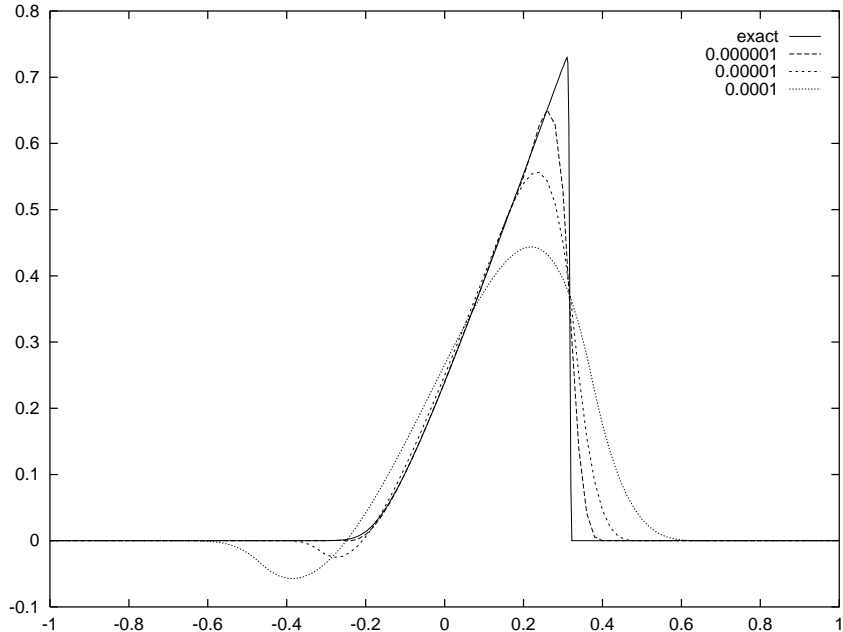}
       }
  }
\end{center}
\caption{{\sf Smooth Dispersion with Exponential Initial Data.  
	      $N=400$,  $T=0.5$, $\Delta t \cdot \delta = 10^{-8}$.  
              The different plots refer to different values of $\delta$
	The ``exact'' curve is the entropy solution of the Burgers
	equation subject to the same initial data} \label{figure:ex1.5}}
\end{figure}

A single ripple which disappears at $\delta$ tends to zero is observed when equation
(\ref{eq:example.abs}) is replaced by the smoother variant (\ref{eq:ex.smooth})
subject to the same exponential initial data.
An example of the approximated solutions of (\ref{eq:ex.smooth})
is presented in figure \ref{figure:ex1.5}.
Once again we observe the appearance of a single dispersive ripple which vanishes
in the limit $\delta \rightarrow 0$.
As before, in that limit, the solution appears to strongly converge to the unique
entropy solution of the Burgers equation subject to the same initial data.

Our numerical results in this section lead us to the following
\begin{conjecture}  \label{conjecture:strong}
  The solution of 
\[
 u_{t}+ f(u)_{x} - \delta g(u_{xx})_{x} = -\epsilon u_{xxxx}, \qquad \delta,\epsilon>0,
\]
with a strictly convex $f$ and a concave $g$, with $g(0)=0$,
strongly converges to the unique
entropy solution of (\ref{eq:scalar.conservation}) as 
$\delta$ and $\epsilon\delta^{-1}$ tend to $0$.
\end{conjecture}

%
%
\section{Study of the Spurious Numerical Oscillations}   \label{section:spurious}
\setcounter{equation}{0}
\setcounter{figure}{0}

In this section we demonstrate and analyze several cases in which numerical oscillations
appear.  
We emphasize that the goal of this study is to gain a better understanding
in order to distinguish between the properties of the model equations and the
phenomena which are strictly related to the numerics in hand.

\begin{figure}[H]    
\begin{center}
  \leavevmode {
  \hbox{
        \epsfxsize=12cm
        \epsffile{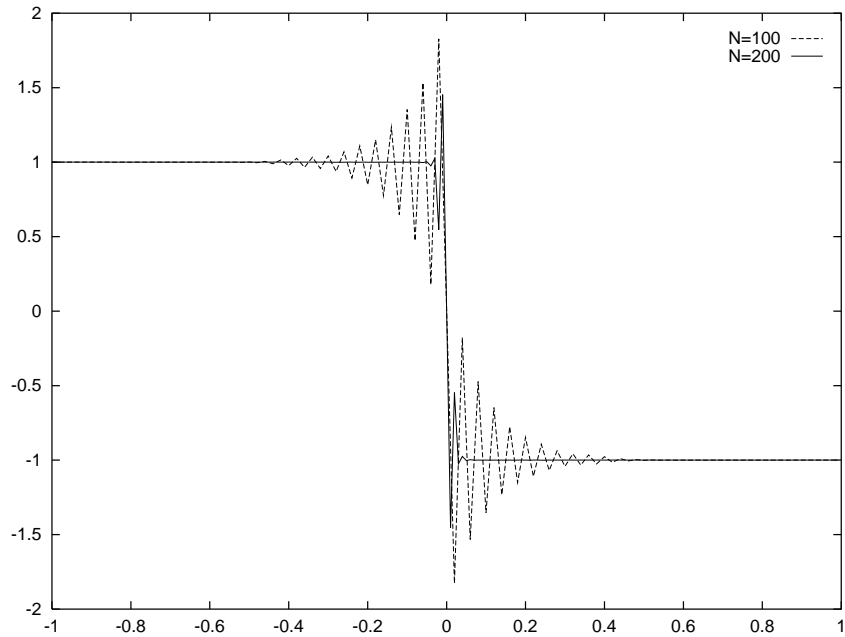}
       }
  }
\end{center}
\caption{{\sf Absolute-Value Dispersion with Riemann Initial Data.
              $\delta=10^{-5}$,  $T=0.5$} \label{figure:ex2.1}}
\end{figure}

\begin{figure}[H]    
\begin{center}
  \leavevmode {
  \hbox{
        \epsfxsize=12cm
        \epsffile{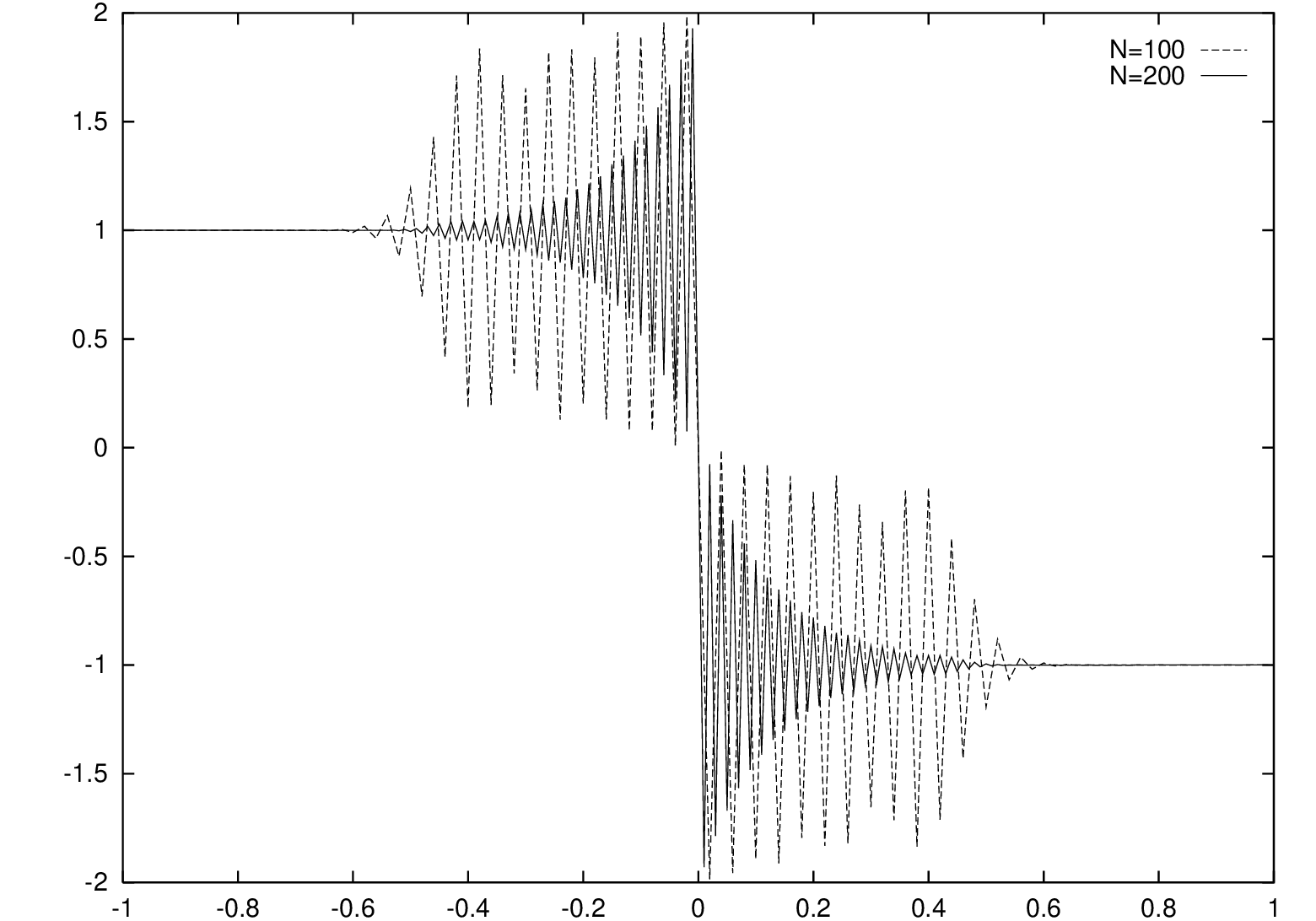}
       }
  }
\end{center}
\caption{{\sf Absolute-Value Dispersion with Riemann Initial Data.
              $\delta=10^{-6}$,  $T=0.5$} \label{figure:ex2.2}}
\end{figure}

\begin{figure}[H]    
\begin{center}
  \leavevmode {
  \hbox{
        \epsfxsize=12cm
        \epsffile{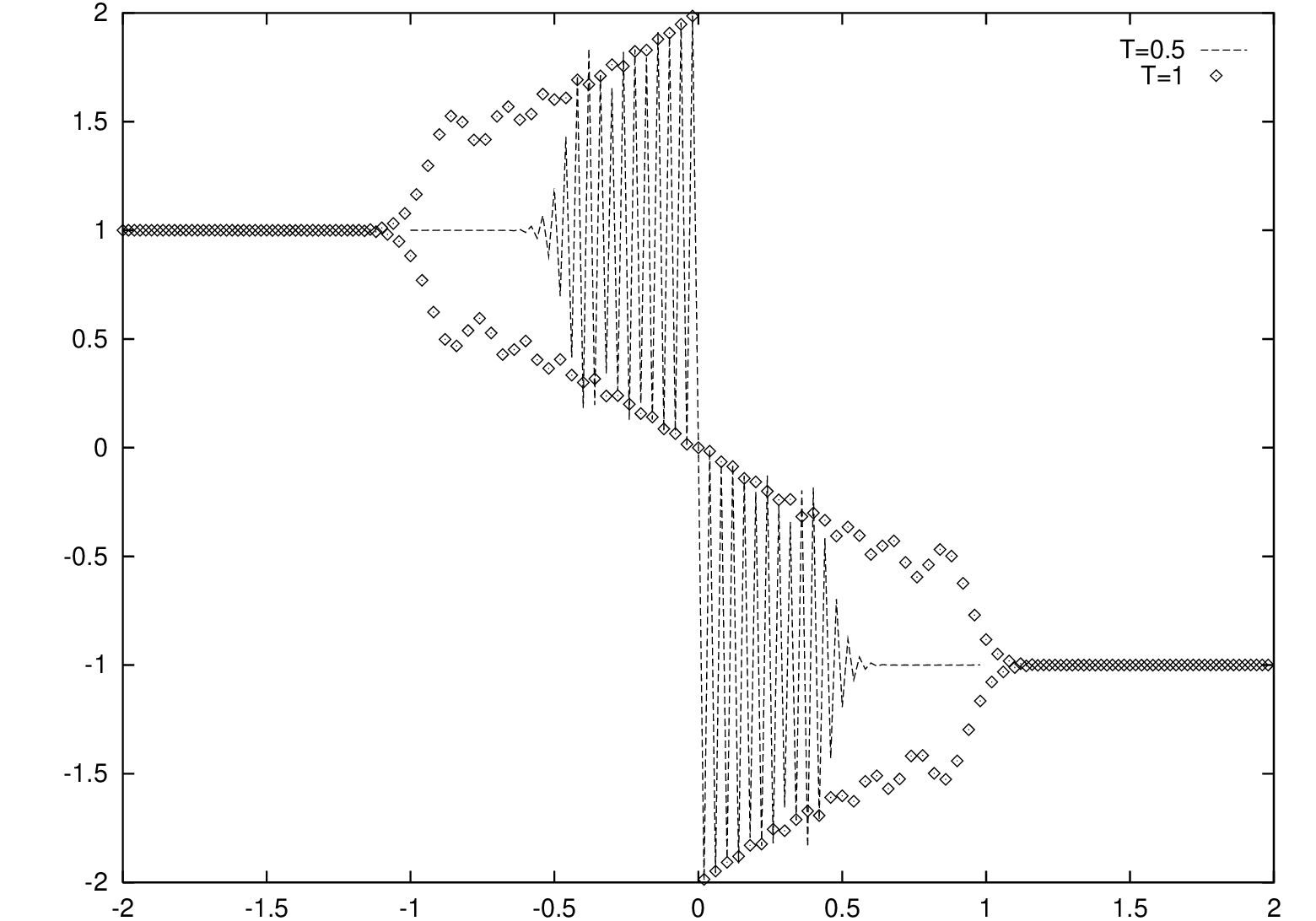}
       }
  }
\end{center}
\caption{{\sf Absolute-Value Dispersion with Riemann Initial Data.
              $\delta=10^{-6}$,  $N=100$} \label{figure:ex2.3}}
\end{figure}

In figures \ref{figure:ex2.1}--\ref{figure:ex2.3} we present results
obtained by approximating solutions of (\ref{eq:mod.2}) with
 $g(u_{xx}) = -|u_{xx}|$,
subject to the Riemann initial data (\ref{eq:example.riemann}) for several 
different sets of parameters.

One can clearly observe the binary oscillations propagating
from the discontinuity at the center of the domain.  These oscillations are strictly
associated with the numerical method.

Note, in particular, the difference between the two oscillatory solutions presented
in figure \ref{figure:ex2.2}.
While the support of the oscillations of the solution for $N=100$ expand in time, the 
solution for $N=200$ is a steady-state.  Even though we do not know what mechanism toggles
between these two states, we are able in the steady-state case to compute the envelope of
the decaying oscillations.

We would like to stress that these oscillations appear in the non-smooth
 $g(s) = -|s|$, but do not appear with the smoother $g(s) =-s^{2}$.  Such
oscillations-free results were already presented in the study of traveling waves
with the smooth $g(s) = -s^{2}$, (Figures \ref{figure:trav.1}--\ref{figure:trav.2}).

\begin{figure}[H]    
\begin{center}
  \leavevmode {
  \hbox{
        \epsfxsize=12cm
        \epsffile{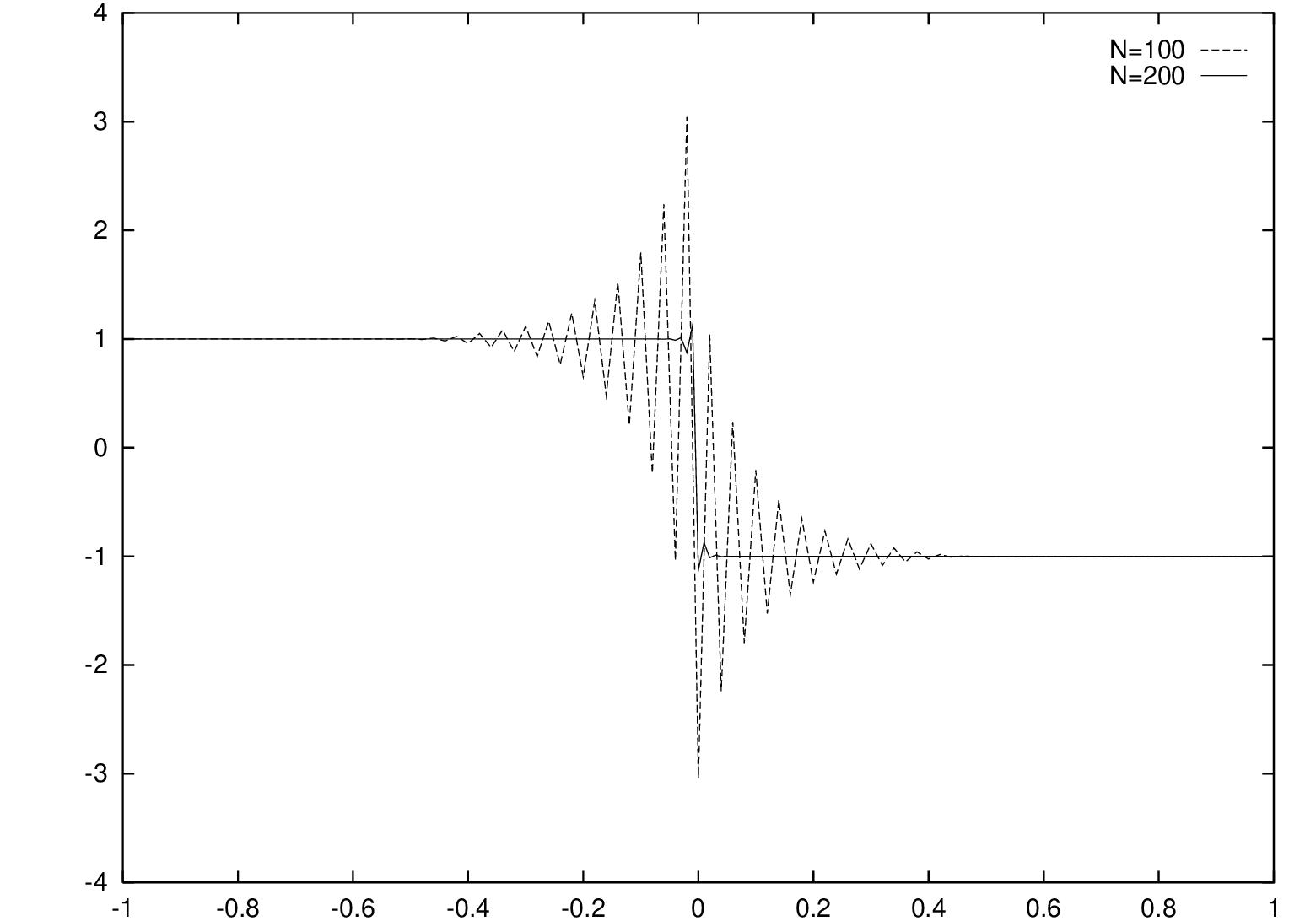}
       }
  }
\end{center}
\caption{{\sf  Absolute-Value Dispersion with Riemann Initial Data which does not include the origin.
               $\delta=10^{-5}$,  $T=0.5$} \label{figure:ex2.a.1}}
\end{figure}

\begin{figure}[H]    
\begin{center}
  \leavevmode {
  \hbox{
        \epsfxsize=12cm
        \epsffile{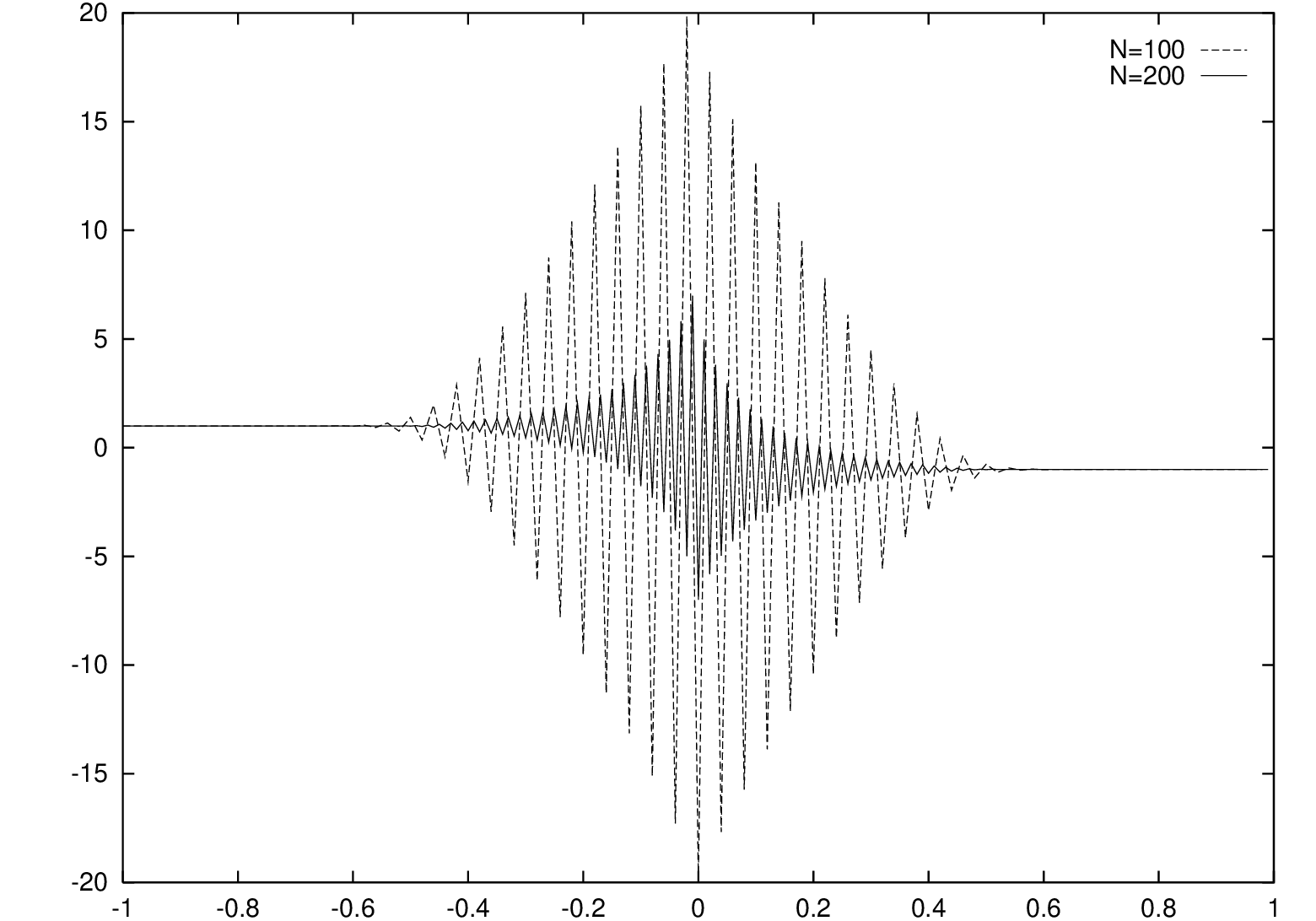}
       }
  }
\end{center}
\caption{{\sf Absolute-Value Dispersion with Riemann Initial Data which does not include the origin.
              $\delta=10^{-6}$,  $T=0.5$} \label{figure:ex2.a.2}}
\end{figure}

\begin{figure}[H]    
\begin{center}
  \leavevmode {
  \hbox{
        \epsfxsize=12cm
        \epsffile{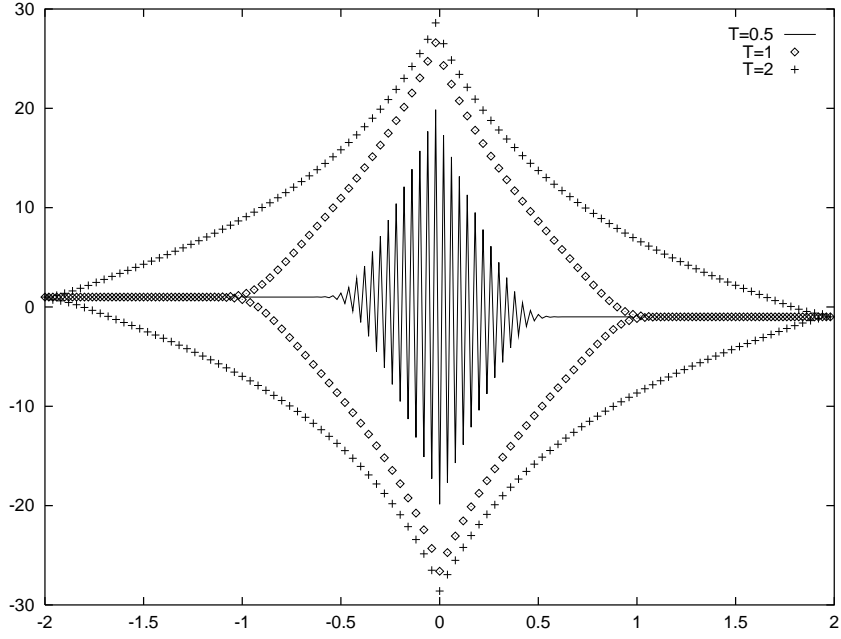}
       }
  }
\end{center}
\caption{{\sf Absolute-Value Dispersion with Riemann Initial Data which does not include the origin.
              $\delta=10^{-6}$,  $N=100$} \label{figure:ex2.a.3}}
\end{figure}

The numerical initial data which was used in figures \ref{figure:ex2.1}--\ref{figure:ex2.3}
includes the origin $(0,0)$.  The existence of this point in the initial data has a strong
stabilizing effect on the results.  
It seems to control the oscillations such that they do not change
sign and they remain bounded in time.
This is not the case when this extra point is not used.  For comparison we bring
in figures \ref{figure:ex2.a.1}--\ref{figure:ex2.a.3} equivalent results to figures
\ref{figure:ex2.1}--\ref{figure:ex2.3} when no such extra point is used.
Similar sensitivity of the oscillations with respect to the location of the grid points
was previously observed in a different context by Hayes in \cite{hayes-97}.

We turn to a study of the binary oscillations presented in the above examples.
The method we use follows the lines of Goodman and Lax in
\cite{goodman-lax:dispersive}.  Similar techniques were used, e.g., 
in \cite{hayes-97, hayes-98, levermore-liu}.

Our starting point is the method of lines (\ref{eq:mod.2.lines}) with the specific
choice of $f(u)=u^{2}/2$.  We assume that the numerical solution introduces 
binary oscillations of the form 
\be
  u_{i} = v_{i} + (-1)^{i} w_{i}.   \label{eq:binary}
\ee
We plug (\ref{eq:binary}) into (\ref{eq:mod.2.lines}).  For the first term,
we have
\bea
 \lefteqn{\frac{1}{\Delta x} \Delta_{-}\left[ f\left( \frac{u_{i}+u_{i+1}}{2} \right) \right] =  
    \frac{1}{8 \Delta x} [(u_{i+1}+u_{i-1}+2u_{i})(u_{i+1}-u_{i-1})] =} \label{eq:bin.f.dis} \\ \nonumber \\
  && = \left\{  \begin{array}{ll}
		\displaystyle{\frac{v_{i+1}-w_{i+1}+v_{i-1}-w_{i-1}+2v_{i}+2w_{i}}{4} 
                  \frac{v_{i+1}-v_{i-1}-w_{i+1}+w_{i-1}}{2\Delta x}} & i \,\, \mbox{even} \\ \\
		\displaystyle{\frac{v_{i+1}+w_{i+1}+v_{i-1}+w_{i-1}+2v_{i}-2w_{i}}{4}
                  \frac{v_{i+1}-v_{i-1}+w_{i+1}-w_{i-1}}{2 \Delta x}} & i \,\, \mbox{odd}.
               \end{array}
      \right. \nonumber
\eea

If we consider equation (\ref{eq:bin.f.dis}) as a discrete approximation of continuous equation,
it can be rewritten as
\be
 \frac{1}{\Delta x} \Delta_{-}\left[ f\left( \frac{u_{i}+u_{i+1}}{2} \right) \right] \longrightarrow
 \left\{  \begin{array}{ll}
     \displaystyle{v(v_{x}-w_{x})} \quad i \,\, \mbox{even} \\
     \displaystyle{v(v_{x}+w_{x})} \quad i \,\, \mbox{odd}.
          \end{array}
 \right.    \label{eq:bin.f.cont}
\ee

For the second term in (\ref{eq:mod.2.lines}) we repeat the same analysis and end with 
\bea
 \lefteqn{\Delta_{-}g(w_{i},w_{i+1}) \longrightarrow} \label{eq:bin.g.cont} \\ \nonumber \\
 && \left\{  \begin{array}{ll}
   \displaystyle{g\left(v_{xx}-\frac{4 w}{(\Delta x)^{2}},v_{xx}+\frac{4 w}{(\Delta x)^{2}}\right) -
         g\left(v_{xx}+\frac{4 w}{(\Delta x)^{2}},v_{xx}-\frac{4 w}{(\Delta x)^{2}}\right)} &
              i \,\, \mbox{even} \\ \\
   \displaystyle{g\left(v_{xx}+\frac{4 w}{(\Delta x)^{2}},v_{xx}-\frac{4 w}{(\Delta x)^{2}}\right) -
         g\left(v_{xx}-\frac{4 w}{(\Delta x)^{2}},v_{xx}+\frac{4 w}{(\Delta x)^{2}}\right)} &
              i \,\, \mbox{odd}.
	  \end{array}
 \right.  \nonumber
\eea
When we specifically choose the flux-limiter as the Engquist-Osher (EO) limiter, 
(\ref{eq:bin.g.cont}) becomes
\be
 \Delta_{-}g(w_{i},w_{i+1}) \longrightarrow 
   \left\{  \begin{array}{ll}
   \displaystyle{sgn(\alpha)g(\alpha) - sgn(\beta)g(\beta)} & i \,\, \mbox{even} \\ \\
   \displaystyle{-sgn(\alpha)g(\alpha) + sgn(\beta)g(\beta)} & i \,\, \mbox{odd},
	  \end{array}
 \right.  \label{eq:bin.g.cont.eo} 
\ee
where
\be
   \alpha = v_{xx}+\frac{4w}{(\Delta x)^{2}}, \quad
   \beta  = v_{xx}-\frac{4w}{(\Delta x)^{2}}.         \label{eq:alpha.beta}
\ee
We require that the continuous solution will satisfy both equation for the even and odd indices.
Combining the results of (\ref{eq:bin.f.cont}) and (\ref{eq:bin.g.cont.eo}), the continuous
equations that correspond to the discretization of the method of lines (\ref{eq:mod.2.lines})
given by the binary oscillations solution (\ref{eq:binary}) are
\be
 \left\{
   \begin{array}{l}
     \displaystyle{(v+w)_{t} + v(v_{x}-w_{x}) - \frac{\delta}{\Delta x}
       (sgn(\alpha)g(\alpha) - sgn(\beta)g(\beta)) = 0}, \\ \\
     \displaystyle{(v-w)_{t} + v(v_{x}+w_{x}) - \frac{\delta}{\Delta x}
       (-sgn(\alpha)g(\alpha) + sgn(\beta)g(\beta)) = 0}, 
   \end{array}
 \right.    \label{eq:bin.cont}
\ee
where $\alpha$ and $\beta$ are given in (\ref{eq:alpha.beta}).   
If we specifically chose
$g(s) = -|s|$ then for $v_{xx} \ll 4w/(\Delta x)^{2}$, (\ref{eq:bin.cont}) becomes
\be
\left\{
   \begin{array}{l}
     \displaystyle{(v+w)_{t} + v(v_{x}-w_{x}) + \frac{8\delta}{(\Delta x)^{3}}w=0}, \\ \\
     \displaystyle{(v-w)_{t} + v(v_{x}+w_{x}) - \frac{8\delta}{(\Delta x)^{3}}w=0}.
   \end{array}
 \right.    \label{eq:bin.cont.abs}
\ee
Here, it is clear that by taking the limit $\delta \rightarrow 0$, in order not to affect the
oscillations that might develop, the ratio $\delta/(\Delta x)^{3}$ has to be held fixed.
Of course, numerically in this case, the RHS has a term $- \epsilon u_{xxxx}$, with 
$\epsilon = \Delta x \delta$ which is of order $\delta / (\Delta x)^{3}$.  Taking $\delta$ large
enough for a fixed $\Delta x$ means that there is more dissipation and in such a way
oscillations can be controlled.

\begin{figure}[H]    
\begin{center}
  \leavevmode {
  \hbox{
        \epsfxsize=12cm
        \epsffile{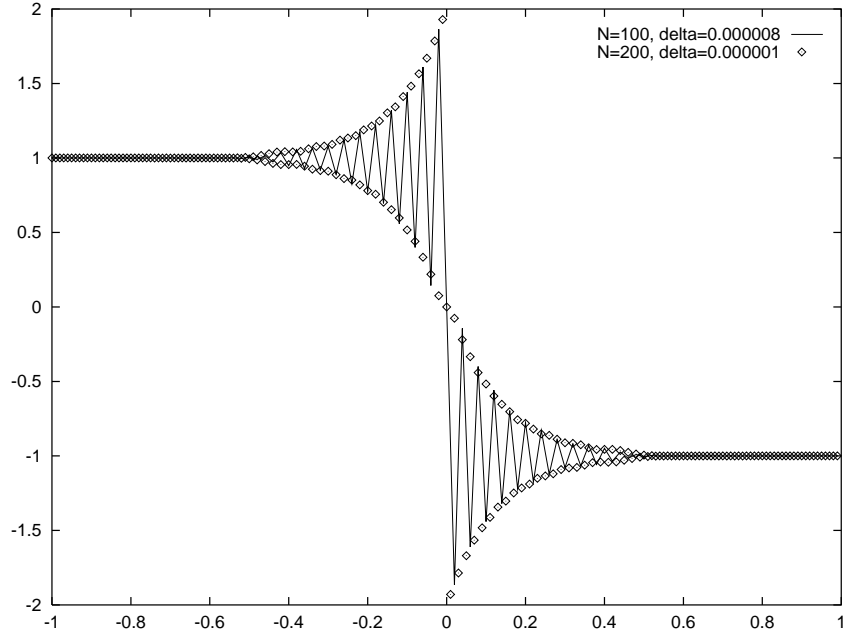}
       }
  }
\end{center}
\caption{{\sf Steady-State Solution, $T=0.5$.
              Fixed $\delta/(\Delta x)^{3}$ Controls the Oscillations} \label{figure:factor.1}}
\end{figure}

An example of such sort of control of the oscillations is shown in figure \ref{figure:factor.1}.
Here, we approximated the solution of (\ref{eq:example.abs}), subject to the Riemann initial
data (\ref{eq:example.riemann}).  The ratio $\delta/(\Delta x)^3$ was held fixed.

As previously noted, in several cases when binary oscillations appear, 
it seems as if there is convergence to steady-state solutions.  
The envelope of this steady-state solutions can be computed from the 
above analysis, since what we are left with in this case amounts to (assuming positive $w>0$)
\be
  \left\{
   \begin{array}{l}
	\displaystyle{v_{t}+v v_{x}=0}, \\ \\
	\displaystyle{w_{t}-v w_{x} + cw=0},
   \end{array}
 \right.    \label{eq:bin.cont.envelope}
\ee
where $c = 8 \delta / (\Delta x)^{3}$.  Then, for $\partial_{t}=0$, we have
\be
  \left\{
   \begin{array}{l}
	\displaystyle{v^{s}v_{x}^{s}=0}, \\ \\
	\displaystyle{-v^{s}w_{x}^{s}+cw^{s}=0}.
   \end{array}
 \right.    \label{eq:bin.cont.steady}
\ee
Hence, $v^{s} = \pm \tilde{c}$ which can be determined by taking the limit $x \rightarrow \pm \infty$.
For the particular choice of $v^{s} = \pm 1$ one has
\[
 w^{s} = \pm \hat{c} e^{\pm c x} = \pm \hat{c} e^{\pm \frac{8 \delta}{(\Delta x)^{3}}x}.
\]
We experimentally observed that when the initial data includes the origin,
the oscillations do not change sign and hence $\hat{c}$ is limited
by the size of the discontinuity.
The envelope of the oscillations is therefore given by 
\[
 v \pm w = \pm 1 \pm e^{\pm \frac{8 \delta}{(\Delta x)^{3}}x},
\]
where the different variations of the sign correspond to the upper/lower envelope at
each part of the solution.

\begin{figure}[H]    
\begin{center}
  \leavevmode {
  \hbox{
        \epsfxsize=12cm
        \epsffile{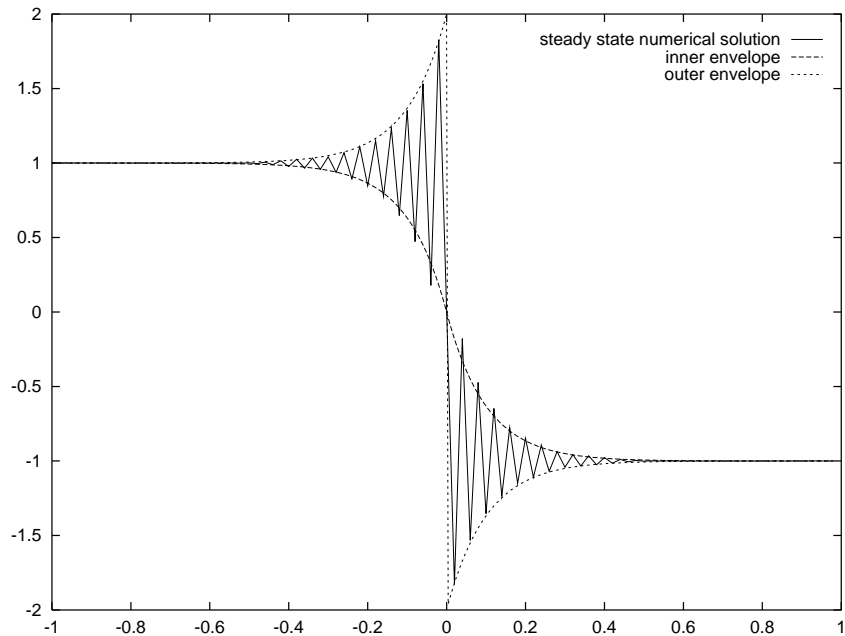}
       }
  }
\end{center}
\caption{{\sf The Envelope of the Steady-State Oscillatory Solution,
              $N=100$, $\delta=10^{-5}$, $T=0.5$} \label{figure:env.1}}
\end{figure}

\begin{figure}[H]    
\begin{center}
  \leavevmode {
  \hbox{
        \epsfxsize=12cm
        \epsffile{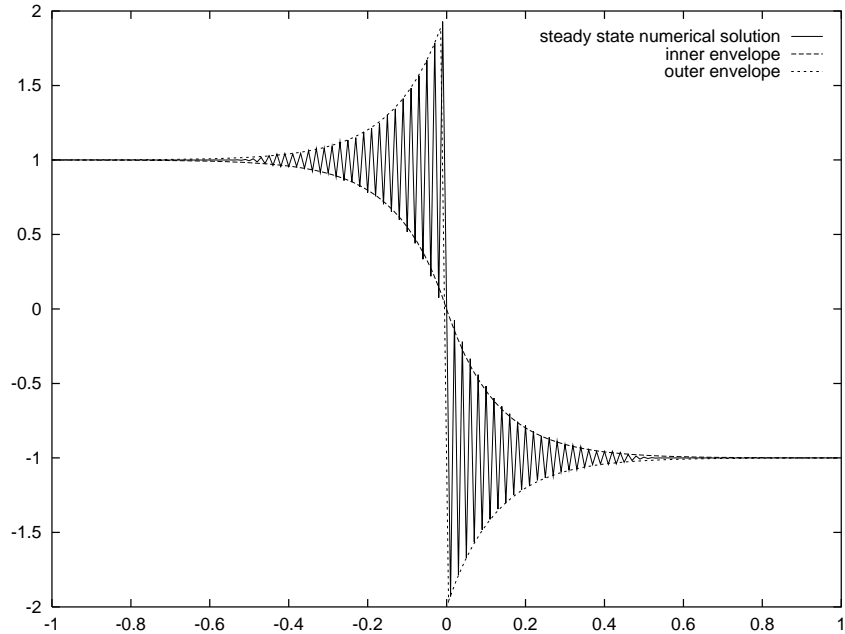}
       }
  }
\end{center}
\caption{{\sf The Envelope of the Steady-State Oscillatory Solution,
              $N=200$, $\delta=10^{-6}$, $T=0.5$} \label{figure:env.2}}
\end{figure}

In figures \ref{figure:env.1}--\ref{figure:env.2} we plot the results obtained by solving 
(\ref{eq:example.abs}) subject to the Riemann initial data (\ref{eq:example.riemann}).  
The numerical solution convergence to an oscillatory steady-state 
solution in which the oscillation are binary.  

In figure \ref{figure:env.1}, the parameters used were $\delta=0.00001$, $N=100$.  The corresponding
outer envelope takes the form
\[
 \left\{
  \begin{array}{ll}
    -1+e^{-c x} & x>0 \\ \\
    1-e^{-c x} & x<0,
  \end{array}
 \right.       \qquad c=10,
\]
while the inner envelope is 
\[
 \left\{
  \begin{array}{ll}
    -1-e^{-c x} & x>0 \\ \\
    1+e^{-c x} & x<0.
  \end{array}  \qquad c = 10,
 \right.
\]

In figure \ref{figure:env.2}, the parameters were changed to $\delta=0.000001$, $N=200$.  
The corresponding outer envelope takes the same form, with a different constant $c=8$.

\begin{thebibliography}{99}

\bibitem{brenier-osher:oslc} Brenier Y., Osher S., 
{\em The Discrete One-Sided Lipschitz Condition for Convex Scalar Conservation Laws}, 
SINUM, {\bf 25}, No. 1 (1988), pp.8-23.

\bibitem{drazing:solitons} Drazin P. G., 
\underline{Solitons}, 
London Math. Soc. Lect. Note Ser. 85, Cambridge university press, 1983.

\bibitem{goodman-lax:dispersive} Goodman J., Lax P. D., 
{\em On Dispersive Difference Schemes, I},
Comm. Pure Appl. Math., {\bf 41} (1988), pp.591-613.

\bibitem{hayes-98} Hayes B. T.,
{\em Binary and Ternary Oscillations in a Cubic Numerical Scheme},
Physica D, {\bf 120} (1998), pp.287-314.

\bibitem{hayes-97} Hayes B. T., 
{\em Binary Modulated Oscillations in a Semi-Discrete Version
of Burgers Equation},
Physica D, {\bf 106} (1997), pp.287-313.

\bibitem{lax-levermore} Lax P. D., Levermore C. D., 
{\em The Small Dispersion Limit of the Korteweg de Vries Equation, III},
Comm. Pure Appl. Math., {\bf 36} (1983), pp.809-829. 

\bibitem{levermore-liu} Levermore C. D., Liu J.-G.,
{\em Large Oscillations Arising in a Dispersive Numerical Scheme},
Physica D, {\bf 99} (1996), pp.191-216.

\bibitem{leveque} LeVeque R. J.,  
\underline{Numerical Methods for Conservation Laws},
Lectures in Mathematics ETH Z\"urich, Birkh\"auser Verlag, Basel, 1992. 

\bibitem{serre} Serre D., 
\underline{Syst\`emes de Lois de Conservation. I},
Diderot Editeur, Paris, 1996.

\bibitem{smoller} Smoller J.,  
\underline{Shock Waves and Reaction-Diffusion Equations},
Springer, 1983.

\bibitem{whitham} Whitham G. B., 
\underline{Linear and Nonlinear Waves}, 
Wiley-Interscience , New York, 1974.



\end{thebibliography}
\end{document}